# Multiple, weak hits confuse complex systems:
# A transcriptional regulatory network as an example

Vilmos Ágoston[1*], Péter Csermely[2] and Sándor Pongor[1,3]

[1]*Szeged Biological Research Center, P. O. Box 521., H-6701 Szeged, Hungary*
[2]*Department of Medical Chemistry, Semmelweis University, P. O. Box 260., H-1444 Budapest 8, Hungary*
[3]*International Centre for Genetic Engineering and Biotechnology, Padriciano 99, I-34012 Trieste, Italy*

Robust systems, like the molecular networks of living cells are often resistant to single hits such as those caused by high-specificity drugs. Here we show that partial weakening of the *Escherichia coli* and *Saccharomyces cerevisiae* transcriptional regulatory networks at a small number (3–5) selected nodes can have a greater impact than the complete elimination of a single selected node. In both cases, the targeted nodes have the greatest possible impact; still the results suggest that in some cases broad specificity compounds or multitarget drug therapies may be more effective than individual high-affinity, high-specificity ones. Multiple but partial attacks mimic well a number of *in vivo* scenarios and may be useful in the efficient modification of other complex systems.

## I. INTRODUCTION

Due to the general applicability of network models [1-3], network damage has become a widely examined phenomenon in various fields. Scale-free networks have been shown to be relatively insensitive to random damage. However, they are rather vulnerable to attacks targeted to their most-connected elements, called hubs [4]. In several networks cascading failures may occur [5-7] and the effects of network topology [4,7-11] permanent damage [12] on the resistance of networks have been examined.

Most of the above studies used a complete elimination of an element from the network to assess network stability. Here we would like to provide a general answer to the following question: Is the partial inactivation of several targets more efficient than the complete inactivation of a single target? The immediate motivation underlying this research is a question relevant to drug design: can broader specificity, lower affinity compounds or multi-drug therapies be more efficient than high affinity, high specificity compounds. The success of multi-target drugs, like Aspirin, Metformin, non-steroidal anti-inflammatory drugs (NSAIDs) and Gleevec™, the recent breakthrough of combinational (cocktail) therapies against AIDS, Alzheimer's disease, asthma, bacterial, fungal and viral infections, diabetes, cancer, hypertension, multiple sclerosis, psychiatric disorders, etc. as well as the efficiency of complex natural remedies (like herbal teas) all suggest that attacking multiple targets may be a useful therapeutic strategy [13-20]. We speculated that this phenomenon might be reflected in the attack vulnerability the biological networks. Using various attack strategies against the *E. coli* [21] and *S. cerevisiae* [22] transcriptional regulatory networks we found that partial weakening at a surprisingly small number of points can be, in fact, more efficient than the complete elimination of a single node. These results prompt further studies to examine the relative efficiency of multitarget drugs and suggest that the examination of multiple attacks can be a promising area for further drug design studies.

## II. METHODS

### A. Networks

We have chosen the regulatory network data of *E. coli* [21] and *S. cerevisiae* [22] as network models. The reason behind this choice was that regulatory proteins provide a plausible framework for modeling drug effects [23,24]. First of all, regulatory mechanisms constitute a very sensitive, central part of the cellular machinery, and their perturbation influences a wide variety of vital functions. Second, regulatory networks belong to a broad class of scale-free networks characteristic of many other biological systems [2]. These networks are directed graphs with 424 nodes/521 edges and 689 nodes/1080 edges, respectively. Loops representing autoregulation were omitted as they do not influence the value of network efficiency (for definition, see Sec. II D). The random networks were generated by distributing the same number of randomly directed edges among the same number of nodes as found in the *E. coli* [21] and *S. cerevisiae* [22] regulatory networks, respectively.

*Email addresses: vilagos@nucleus.szbk.u-szeged.hu;
csermely@puskin.sote.hu; pongor@icgeb.org

### B. Attack strategies

The attack of a single target was performed by the elimination of all interactions at the representing node [Fig. 1(A), *complete knockout*]. Partial inactivation of a target was modeled in two different ways. Either half of the interactions of a given node [Fig. 1(B1), *partial knockout*] has been removed, or all interactions of the element were attenuated [Fig. 1(B2), *attenuation*; for the description of attenuation see Sec. II E]. Finally, a distributed, system wide attack can affect any protein-protein interaction (any edge) within the network. Again, we used two simplified strategies, knockout [Fig. 1(C1), *distributed knockout*] or attenuation of individual interactions (edges) of the network [Fig. 1(C2), *distributed attenuation*; see Sec. II E].

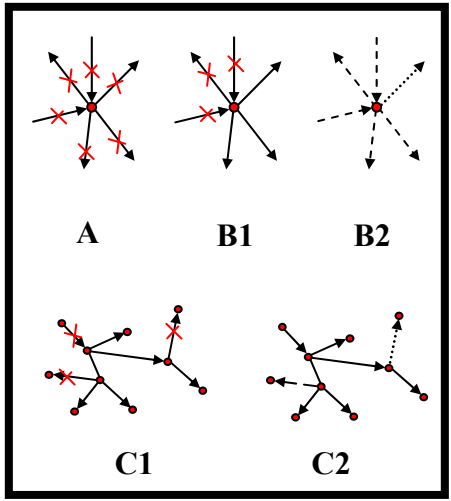

FIG. 1. (Color) Partial target inhibition strategies. Panel (A), complete knockout: complete inhibition of a single target modeled by the elimination of all interactions at the representing node. Panel (B1), partial knockout: partial inactivation of the target by knocking out half of its interactions. Panel (B2), attenuation: partial inactivation of the target by attenuating the interactions of the representing node to 50% as an average. Panel (C1), distributed knockout: inactivation of individual interactions between nodes. Panel (C2), distributed attenuation: attenuation of individual interactions between nodes. In the attenuation experiments, attacking an edge at one end resulted in a 50% weakening of the interaction (dashed line). If a subsequent attack is directed against the other end of the edge, the interaction is weakened to 25% (dotted line).

We can translate these models into biochemical terms by saying that a high affinity drug can knock out an interaction, while a low affinity drug will only attenuate it. Similarly, a highly specific drug is able to target one single interaction, while less specific drugs will affect more/all interactions of a given node (protein or operon). Needless to say, there is a multitude of other possibilities. Those above were chosen only as characteristic examples in order to test whether a combination of several partial inactivation events can reach an effect at least equivalent to the knockout of a single target.

### C. Successive maximal damage strategy

Simulation experiments were based on a successive maximal damage strategy. The search for maximal damage caused by multiple attacks is difficult in a combinatorial context. For instance, if we want to determine which 5 of the 1000 edges of a given network need to be deleted in order to produce a maximal effect on the network efficiency (NE, see part II.D.), we would need to test $1000!/(5!*995!) > 8*10^{11}$ cases in a single simulation experiment. Instead, we used a greedy algorithm by choosing the elements whose step-by-step removal produces the largest damage. This was carried out by first determining the damage caused by the removal of each individual node (or edge, depending on the strategy; see Fig. 1). The node or edge causing the maximum damage was selected for removal in the subsequent attack. In the above example, this procedure leads to a quasi optimal solution in less than 5000 steps. We have to note, that the network efficiency value obtained in this manner is only an upper estimate of the maximal damage, since there may be more efficient combinations.

### D. Network efficiency

The damage induced by the attacks on the networks was monitored by calculating their network efficiency (NE). The NE of a simple (undirected, unweighted) graph of *n* nodes is expressed as

$$NE = \sum_{i \neq j} \frac{1}{d_{ij}},$$

where $d_{ij}$ is the shortest path between nodes *i* and j [25]. If the network is directed, $d_{ij}$ is the shortest directed path, if it is weighted, $d_{ij}$ is the path with a minimum weight. Usually, this quantity is divided by the corresponding sum of a fully connected network to give a relative network efficiency between 0 and 1. In our case this was not necessary, since we used the network efficiency of the starting network as 100%. The decrease of NE was plotted as a function of the attacks.

### E. Attenuation experiments

In the attenuation experiments, the initial network was unweighted and an attack to an edge was modelled by doubling its weight from 1 to 2.

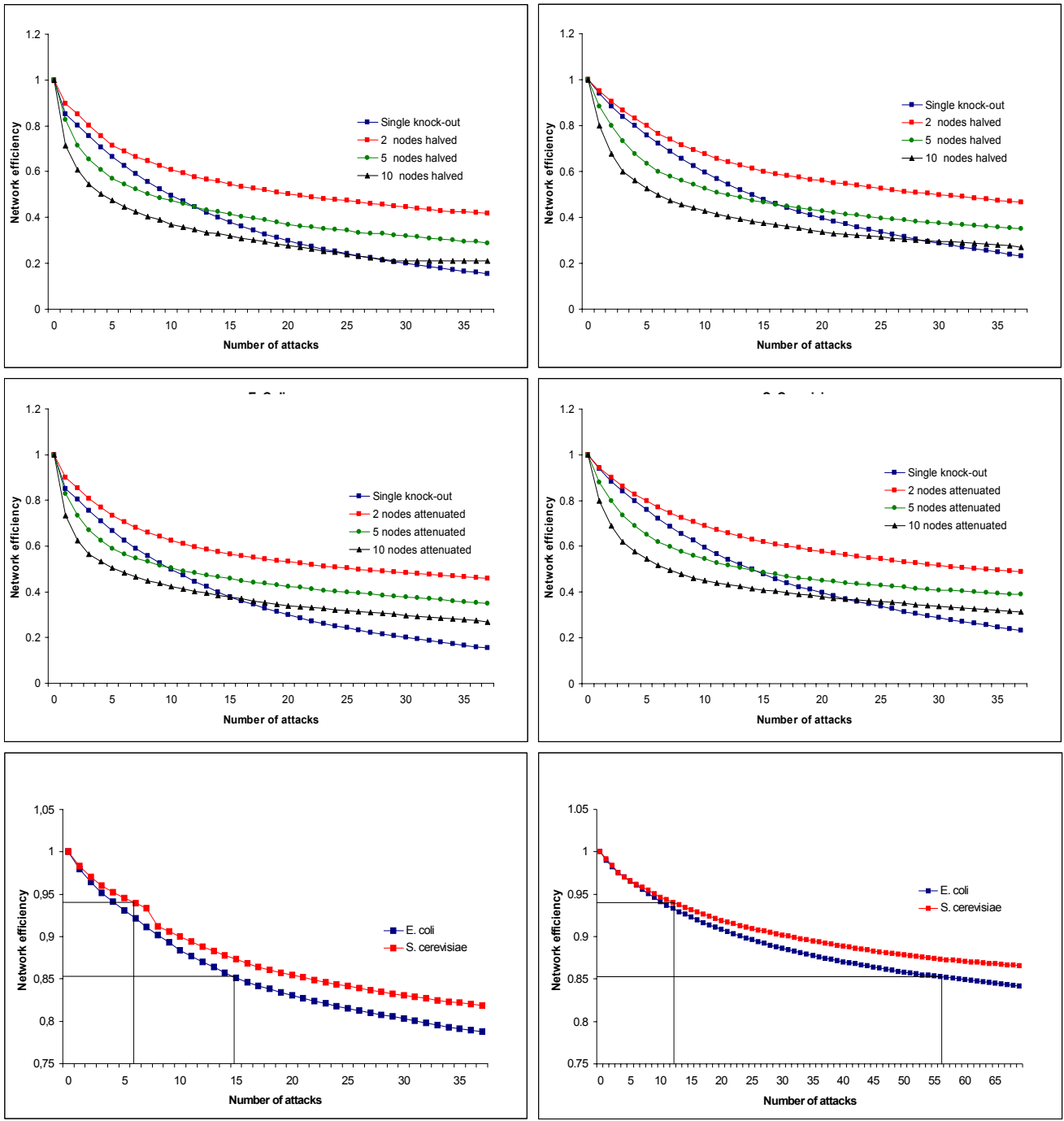

FIG. 2. (Color) Effect of single-target and various multitarget attack strategies on network efficiency. The effect of a series of successive attacks is shown on the network efficiency (NE [25]; see Methods) of the regulatory networks of *E. coli* [21] or *S. cerevisiae* [22]. Each attack point was chosen to produce the maximal possible damage to the system. Panels (A) and (B), single-target attack was performed by eliminating all the edges of a given node [blue; cf. Fig. 1(A)]; partial knockout was modelled by fully blocking (removing) a randomly chosen half of the edges belonging to a given node as shown in Fig. 1, panel (B1). This attack was applied simultaneously to 2 (red), 5 (green) and 10 (black) nodes. Panels (C) and (D), attenuation was modelled by decreasing the contribution of edges belonging to a given node as shown in Fig. 1, panel (B2). The colour codes are the same as in panels (A) and (B). Distributed system-wide knockout was modelled by either fully blocking [removing, Panel (E)] or attenuation [Panel (F)] an edge so as to produce a maximum decrease in NE, as schematically shown in Fig. 1, panels (C1) and (C2), respectively. In the attenuation experiments an edge could be attenuated at both ends, i.e. the maximal attenuation of a single edge was four fold (from the initial 100% to 25%). For this reason the number of attacks [panel (F)] and the number of edges affected (Table I, column 12) do not necessarily coincide. Blue and red signs of panels (C) and (D) refer to data from *E. coli* and *S. cerevisiae*, respectively.

In the calculation of network efficiency, the weight of the shortest path $d_{ij}$ was taken as equal to the highest weight within the path. This means that an attenuated edge within a path was considered to diminish the contribution of the entire path in a bottleneck fashion. Each edge could be attacked at both ends to reach a maximal weight of 4. In each step the node (or edge) to be attacked by attenuation was chosen on the basis of maximal damage.

### III. RESULTS AND DISCUSSION

#### A. Comparison of complete and partial knockouts

To answer the question: is partial inactivation of several targets is more efficient than the complete inactivation of a single target, we used the *E. coli* [21] and S. *cerevisiae* [22] network models described in Sec. II A. Using of the various attack strategies of Sec. II B. the network becomes less connected, and routes between distant nodes become more complicated [4]. It is worth mentioning that removal of the most connected nodes did not always imply the maximal damage of the regulatory networks (directed, weighted graphs) studied here. For instance, in the *S. cerevisiae* network the maximal damage is caused by the removal of the GCN4 node, which has 18 edges, whereas the STE12 node has 71 edges. This is in agreement with the earlier results of Latora and Marchiori [26], who showed that the damage of the most connected nodes is not always the worst damage of the network. The importance of other parameters than the degree of the affected node in determining network damage, like betweenness centrality was also described in other publications [27,28]. These findings were one of the reasons why we performed a rigorous search rather than simply attacking the next most connected node.

The descending curves of Figs. 2(A) and 2(B) show that the complete knockout of single nodes (blue) is more effective than the attenuation of all interactions of two nodes (red). On the other hand, an attenuation of 5 nodes (green) is already more effective than the complete inactivation of a single-target (blue). The same result was found both in the *E. coli* and in the *S. cerevisiae* networks. The effect of attenuation of all interactions at a given node [Figs. 2(C) and 2(D)] proved to be rather similar. Attenuation of 5 nodes (green) produced roughly the same effect as the complete inactivation of a single node (blue). The effect of the third strategy, the distributed system-wide attack is directed against edges, rather than nodes, so the graphic comparison [Figs. 2(E) and 2(F)] is different from the previous cases. It is apparent, however, that the effect produced by the complete elimination of the first node and its 72 edges in the *E. coli* network [Fig. 2(A), first point of the blue line] is reached by the knockout of 15 edges only [see the corresponding value on panel (E)]. Similarly, the complete elimination of the first node and its 18 edges in the *S. cerevisiae* network [Fig. 2(B), first point of the blue line] is reached by the knockout of 6 edges only [see the corresponding value on panel (E)]. The distributed attenuation strategy [Fig. 2(F)] is less efficient, since here 56 or 13 attenuation steps have to be performed in the *E. coli* or *S. cerevisiae* networks, respectively, to achieve the same effect. We note that the simulations shown here are inhibition scenarios, where functions are entirely or partially blocked similar to what happens when an antibiotic acts on a pathogen. The effect of a therapeutic agent that restores the normal function of an inhibited receptor can be modeled by analogous steps carried out in a reverse order.

Turning back to the context of drug design, we attempted a more detailed comparative analysis of the damage after the inactivation of a single node, which is a better analogy to high-affinity, single-target drug-induced effects than the successive maximal damage strategy of Fig. 2. Here our main question was: How many partial attacks are equivalent to the complete inactivation of a single node? A detailed quantitative comparison is shown in Table I. The data represent the number or nodes/edges that have to be attacked by various strategies to produce the same effect (maximal damage) on network efficiency as that of the complete knockout of a single node. In particular, one is tempted to think that multi-target attacks may affect more edges to obtain the same effect as single target knockout, but the results show that this is not necessarily the case. In the *E. coli* network, the *partial knockout* of about 4 nodes is necessary to produce the same effect as the complete elimination of a single node. A total of about 65 edges are deleted in this way, in contrast to the 72 edges of the single eliminated target. *Attenuation* is less efficient; there, 5 nodes and 129 edges have to be attacked in order to reach the same effect. *Distributed knockout* is the most efficient in this respect. As noted above, the elimination of 15 edges of the *E. coli* or 6 edges of the *S. cerevisiae* networks produce the same effect as the elimination of a single node with its 72 or 18 edges, respectively, in these networks. In both the *E. coli* and *S. cerevisiae* networks the fully damaged single node was among the 3-5 partially affected nodes (see footnotes of Table I). Distributed attenuation was less efficient than distributed knockout, especially in terms of the number of edges that had to be attacked in order to reach the same damage. Even though attenuation strategies (corresponding to low affinity drugs) were found less efficient in these calculations than the corresponding knockout strategies (high affinity binders) a slight increase in the number of targets can easily compensate for this disadvantage.

Table I Quantitative comparison of single-target knockout with various multi-target attack strategies.

| Network | (A) Single target Knockout | | | (B) Partial inactivation of several targets | | | | C) Distributed system-wide attack | | | |
|---|---|---|---|---|---|---|---|---|---|---|---|
| | | | | (B1) Partial knockout: half of edges deleted[m] | | (B2) Attenuation of all edges | | (C1) Distributed knockout of individual edges | | (C2) Distributed attenuation of individual edges | |
| | # of nodes deleted | # of edges affected | Damage (% decrease in NE) | Equivalent # of nodes | # of edges affected | Equivalent # of nodes | # of edges affected | Equivalent # of edges affected | # of nodes affected (% of edges)[a] | Equivalent # of edges affected | # of nodes affected (% of edges)[a] |
| 1 | 2 | 3 | 4 | 5 | 6 | 7 | 8 | 9 | 10 | 11 | 12 |
| *E. coli* regulatory network (N=424, E=521) | 1 | 72[b] | 15% | 4.2 | 64.8[c] | 5 | 129[d] | 15 | 19 (5.8%)[e] | 38[f] | 53 (10.5%)[g] |
| *S. cerevisiae* regulatory network (N=689, E=1080) | 1 | 18[h] | 6% | 2.8 | 61.0[i] | 3 | 142[j] | 6 | 11 (3.1%)[f] | 10[l] | 16 (5.4%)[l] |
| Random directed network (N=424, E=521)[m] | 1 | 6.0 | 20% | 2.0 | 5.8 | 4.0 | 19.4 | 2.0 | 4.0 (19.7%) | 5.0 | 8.2 (10.24%) |
| Random, directed network (N=689, E=1080)[m] | 1 | 8.2 | 7% | 2.0 | 6.4 | 2.0 | 7.6 | 2.0 | 4.0 (10.1%) | 3.0 | 6.0 (9.84%) |

[a] E.g. the 15 edges attacked in the *E. coli* network represent 5.8% of the total of 328 edges that belong to the 19 nodes affected by the attack. (In this particular case 11 nodes of the maximal possible 30 affected nodes were overlapping at the different edges.)

[b] Affected operons (# of edges): crp(72)

[c] Affected operons(# of edges): crp(72), rpoH (14), fliAZY (14), fnr (22), arcA (21), rpoE_rseABC (24)

[d] Affected operons (# of edges): crp (72), rpoH (14), fnr (22), fliAZY (14), flhDC (10)

[e] Affected operons (# of edges):: arcA (21), cpxAR(10), crp (72), cspA (2), cytR (7), dnaA (2), flhDC (10), fliAZY (14), fnr (22), fur (10), hns (8), malt (7), mlc (4), nlpD_rpoS (14), ompR_envZ (7), rpoE_rseABC (24), rpoH (14), soxR (1), soxS (7)

[f] The number of attacks (e.g.: 56) can be higher than the number of edges attacked (e.g.: 38) since each edge could be attacked twice. See (*17*) and the legend to Fig. 2.

[g] Affected operons (# of edges): arcA (21), cpxAR(10), crp (72), cspA (2), cytR (7), dnaA (2), flhDC (10), fliAZY (14), fnr (22), fur (10), hns (8), malt (7), mlc (4), nlpD_rpoS (14), ompR_envZ (7), rpoE_rseABC (24), rpoH (14), soxR (1), soxS (7), acrAB (1), acrR (1), ada_alkB (2), adiA (1), adiA_adiY (1), aidB (3), alkA (2), appCBA (2), appY (3), atoC (3), betIBA (2), caiF (6), caiTABCDE (3), exuR (3), fadR (5), fecABCDE (1), fecIR (2), fhlA (4), fixABCX (2), fpr (2), GalR (2), gals (3), glnALG (4), himA (21), hypABCDE (3), iclMR (3), marRAB (6), metJ (4), metR (4), nac (4), nagBACD (4), rpoN (13), rtcR (2), uxuABR (2)

[h] Affected operons (# of edges): IME1 (18)

[i] Affected operons (# of edges): IME1 (18), STE12 (71), GCN4 (53)

[j] Affected operons (# of edges): IME1 (18), STE12 (71), GCN4 (53)

[k] Affected operons (# of edges): SNF2_SWI1 (20), SIN3 (13) SWI5 (11), MCM1 (13), HAP2_3_4_5 (26), MIG1 (26), DAL80 (20), DAL80_GZF3 (5), GAT1 (6), HSF1 (15), UME6 (38)

[l] Affected operons (# of edges): SNF2_SWI1 (20), SIN3 (13) IME1 (18), RME1 (8), IME1_UME6 (4), HAP2_3_4_5 (26), MIG1 (26), SWI5 (11), MCM1 (13), DAL80 (20), DAL80_GZF3 (5), GAT1 (6), HSF1 (15), UME6 (38), GAL4 (14), IME4 (2)

[m] The results are the average of 10 simulations, hence the resulting numbers are not necessarily integers.

Table II. Damage caused by different strategies upon removal of the same number of edges.

| Network | (A) Single target knockout | | | Damage (% decrease in NE) caused by removing the same # of edges | | | |
|---|---|---|---|---|---|---|---|
| | | | | (B) Partial inactivation of several targets | | (C) Distributed system-wide attack | |
| | # of nodes deleted | # of edges affected | Damage (% decr. in NE) | (B1) Partial KO | (B2) Att. of all edges | (C1) Distributed knockout | (C2) Distributed attenuation |
| **1** | **2** | **3** | **4** | **5** | **6** | **7** | **8** |
| *E. coli* regulatory network | 1 | 72 | 15% | 19.9% | 7.4% | 26.9% | 16.9% |
| *S. cerevisiae* regulatory network | 1 | 18 | 6% | 3.4% | 3.0% | 14.0% | 7.6% |

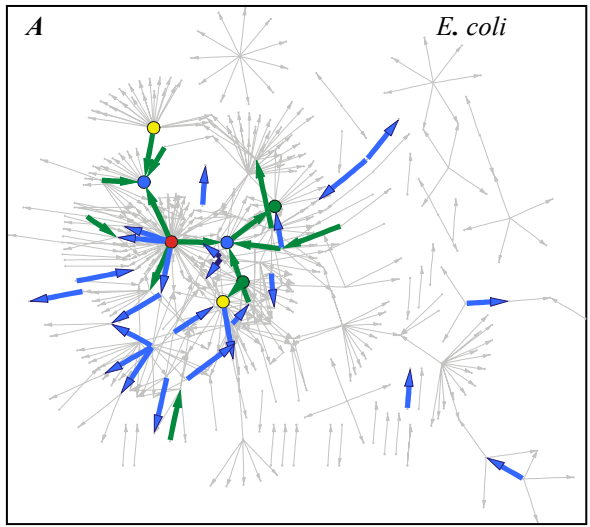
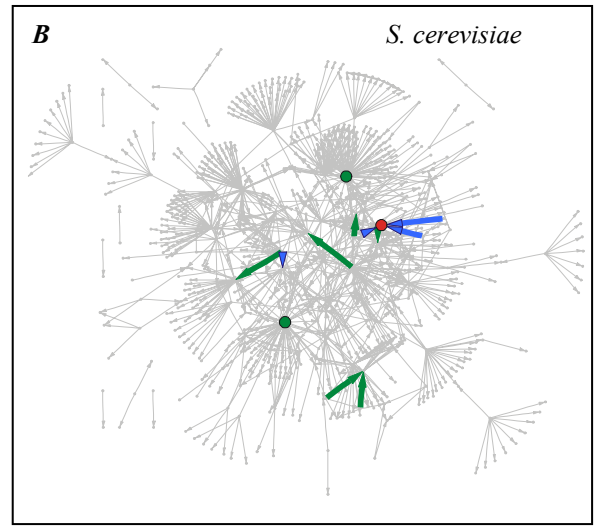

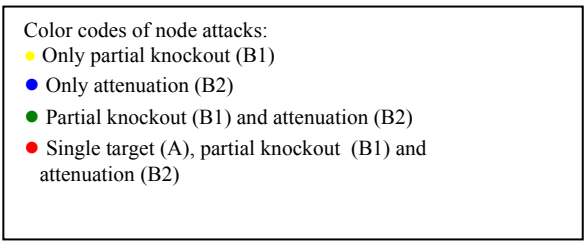
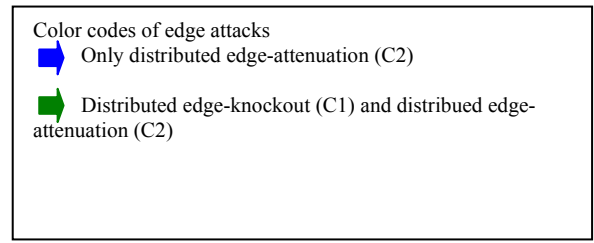

Color codes of node attacks:
- Only partial knockout (B1)
- Only attenuation (B2)
- Partial knockout (B1) and attenuation (B2)
- Single target (A), partial knockout (B1) and attenuation (B2)

Color codes of edge attacks
- Only distributed edge-attenuation (C2)
- Distributed edge-knockout (C1) and distribued edge-attenuation (C2)

FIG. 3. (Color) Sites affected by the various strategies in the *E. coli* (A) and the *S. cerevisiae* (B) regulatory networks. The attacks were carried out with the maximum damage algorithm based on the rigorous search strategy described in Methods. The strategies are those defined in Fig. 1, and the nodes/edges are the same as those described in Table II.

### B. Sites of attacks

Figure 3 shows the sites of the various attacks quantified in Table I in the *E. coli* [Fig. 3(A)] and *S. cerevisiae* [Fig. 3(B)] networks. All strategies target a central, connected part in both networks. On the other hand, in the *S. cerevisiae* network [Fig. 3(B)] the majority of the edges selected by the edge-directed strategies [(C1), (C2) of Fig. 1] are not directly connected to the nodes targeted to by the node-directed strategies [(B1), (B2) of Fig. 1], while most of the attacked edges are connected or close to the attacked nodes in the *E. coli* network [Fig. 3(A)].

### C. Single and multiple hits on random networks

As a comparison, the same attack strategies were applied to random networks [29,30] that have the same number of nodes and edges as the *E. coli* and *S. cerevisiae* regulatory networks, respectively. These random networks also show a rather high susceptibility to multiple, although partial, hits, if compared to the deletion of their single nodes. Moreover, random networks seem to be more susceptible to multi-target attacks than their natural counterparts, since the attack of fewer nodes and fewer edges produces the same damage as in the *E. coli* and *S. cerevisiae* regulatory networks. For example, if one compares the extent of damage (Table I, column 4) and the number of edges necessary for distributed knockout (Table I, column 9), one can see that the elimination of one edge results in about 1% damage in both the *E. coli* and in the *S. cerevisiae* regulatory networks, while in the corresponding random networks the elimination of a single edge corresponds to 10% and 3.5% damage, respectively. We are aware of the fact that the comparison of *E. coli* and *S. cerevisiae* regulatory networks with the corresponding random networks may not be generalized to networks with other topologies, nevertheless, we feel that it is safe to conclude that the susceptibility of networks to multi-target attacks may depend on their topology. In the present two cases we found that the natural, directed networks are somewhat more robust against multi-target attacks than their random counterparts. However, the general validity of this conclusion needs a more thorough analysis.

### D. Multiple hits remain more efficient even if the same number of edges is removed

As mentioned above, the number of eliminated/attenuated edges differed in the various attacks on the *E. coli* and *S. cerevisiae* regulatory networks. This raises the concern that the difference between the various attack-strategies is caused by the unequal number of damaged, removed or partially blocked edges. In Table II we show a comparison where the damage in network efficiency was calculated with an equalized number of deleted edges in each attack scenario. This data confirms that most of the multiple-target strategies shown here can be more efficient than the knockout of a single target, even when the damage of only an equal number of edges is permitted. In the case of the *E. coli* network 3 out of the 4 multiple-target strategies were more efficient than single target knockout, while in the case of the *S. cerevisiae* network half of them were more efficient. The efficiency of multi-target attacks is not trivial: they are not only better because they affect the network in more sites. They can, especially if distributed in the entire network, confuse complex systems more than concentrated attacks even if the number of targeted interactions is the same.

### III. SUMMARY AND CONCLUSIONS

In summary we can conclude that the efficacy of multi-target attacks compares well with that of single-target knockout. Partial knockout or attenuation of a surprisingly small number of targets (e.g. 3 or 5) may produce a larger effect than the complete knockout of a single target. Our studies suggest that certain drugs with multiple targets [13-15] or carefully designed drug combinations [17-20] might have a better chance to affect the complex equilibrium of the whole system than single target drugs [16]. Moreover, it is sufficient that these multi-target drugs affect their targets only partially, which corresponds well with the presumed low-affinity interactions of these drugs with several of their targets [16,31,32]. It has been summarized before that weak links (low-probability, low-intensity edges) stabilize complex systems [33,34]. Here we showed a kind of reverse statement: that multiple, weak hits efficiently confuse the integrity of complex systems. Since the increased sensitivity to small but multiple hits versus major single hits was found in two quite different network types (characterized by scale-free and random topologies, respectively) it may be worthwhile to test this phenomenon in the case of network representations used in areas other than genetic regulatory networks [1-3]. Partial attacks mimic well the physiological scenarios, where a complete elimination of a network node is a rather unusual phenomenon. The partial attack strategy might be worth to try in other models, like the selective removal of nodes and edges to restrict the damage of cascading overload failures [35].

### ACKNOWLEDGEMENTS

Work in the authors' laboratory was supported by research grants from the EU (FP6506850), Hungarian Science Foundation (OTKA-T37357), Hungarian

[1] D.J. Watts and S.H. Strogatz, Nature **393**, 440 (1998).
[2] A.L. Barabasi and R. Albert, Science **286**, 509 (1999).
[3] M.E.J. Newman, SIAM Rev. **45**, 167 (2003).
[4] R. Albert, H. Jeong, and A.L. Barabasi, Nature **406**, 378 (2000).
[5] P. Bak, C. Tang and K. Wiesenfeld, Phys. Rev. Lett. **59**, 381 (1987).
[6] Y. Moreno, J.B. Gomez and A.F. Pacheco, Europhys. Lett. **58**, 630 (2002).
[7] D.J. Watts, Proc. Natl. Acad. Sci. U. S. A. **99**, 5766 (2002).
[8] M.E.J. Newman, Phys. Rev. E **67**, 026126 (2003).
[9] B. Shargel, H. Sayama, I.R. Epstein and Y. Bar-Yam, Phys. Rev. Lett. **90**, 068701 (2003).
[10] G. Paul, T. Tanizawa, S. Havlin and H.E. Stanley, Eur. J. Phys. B **38**, 187 (2004).
[11] A.X.C.N. Valente, A. Sarkar and H.A. Stone, Phys. Rev. Lett. **92**, 118702 (2004).
[12] S.N. Dorogovtsev and J.F.F. Mendes, Phys. Rev. E **63**, 056125 (2001).
[13] S. Huang, Drug Discov. Today **7**, S163 (2002).
[14] W.G.Jr. Kaelin, Science STKE **225** pe12 (2004).
[15] A.A. Borisy, P.J. Elliott, N.W. Hurst, M.S. Lee, J. Lehár, E.R. Price, G. Serbedzija, G.R. Zimmermann, M.A. Foley, B.R. Stockwell and C.T. Keith, Proc. Natl. Acad. Sci. U. S. A. **100**, 7977 (2003).
[16] P. Csermely, V. Ágoston and S. Pongor, Trends Pharmacol. Sci. **26,** 178 (www.arxiv.org/abs/q-bio.BM/0412045) (2005).
[17] B. Schmitt, T. Bernhardt, H.J. Moeller, I. Heuser and L. Frolich, CNS Drugs **18**, 827 (2004).
[18] B. Weinstock-Guttman and R. Bakshi, CNS Drugs **18**, 777 (2004).
[19] I. Gavras and T. Rosenthal, Curr. Hypertens. Rep. **6**, 267 (2004).
[20] S. Del Prato and L. Volpe, Expert Opin. Pharmacother. **5**, 1411 (2004).
[21] S.S. Shen-Orr, R. Milo, S. Mangan, and U. Alon, Nature Genet. **31**, 64 (2002).
[22] R. Milo, S. Shen-Orr, S. Itzkovitz, N. Kashtan, D. Chklovskii and U. Alon, Science **298**, 824 (2002).
[23] B.N. Kholodenko, A. Kiyatkin, F.J. Bruggeman, E. Sontag, H.V. Westerhoff and J.B. Hoek, Proc. Natl. Acad. Sci. U. S. A. **99**, 12841 (2002).
[24] J. Tegnér, M.K.S. Yeung, J. Hasty and J.J. Collins, Proc. Natl. Acad. Sci. U. S. A. **100**, 5944 (2003).
[25] V. Latora, and M. Marchiori, Phys. Rev. Lett. **87**, 198701 (2001).
[26] V. Latora, and M. Marchiori, Phys. Rev. E **71,** 015103 www.arxiv.org/abs/cond-mat/0407491 (2005).
[27] P. Holme, B.J. Kim, C.N. Yoon and S.K. Han, Phys. Rev. E **65**, 056109 (2002).
[28] R. Kinney, P. Crucitti, R. Albert and V. Latora, www.arxiv.org/abs/cond-mat/0410318 (2004).
[29] P. Erdős and A. Rényi, Publicationes Mathematicae Debrecen **6**, 290 (1959).
[30] P. Erdős and A. Rényi, Magyar Tud. Akad. Mat. Kutató Int. Közl. **5**, 17 (1960).
[31] S.A. Lipton, Nature **428** 473 (2004).
[32] M.A. Rogawski, Amino Acids **19** 133 (2000).
[33] P. Csermely, Trends Biochem. Sci. **29**, 331 (2004).
[34] I.A. Kovacs, M.S. Szalay and P. Csermely, FEBS Lett. **579,** 2254 www.arxiv.org/abs/q-bio.BM/0409030 (2005).
[35] A.E. Motter, Phys. Rev. Lett. 93, 098701 (2004).